\documentclass[11pt,a4paper]{article}
\begin{document}

\title{ The effect of the tortoise coordinate on the stable study of
 the Schwarzschild black hole  \footnote{ E-mail of Tian:
 hua2007@126.com, tgh-2000@263.net}}
\author{Tian Gui-hua,\ \ Shi-kun Wang,  \ \ Shuquan Zhong\\
School of Science, Beijing University \\
of Posts And Telecommunications. Beijing100876, China.\\Academy of
Mathematics and Systems Science,\\ Chinese Academy of
Sciences,(CAS) Beijing, 100080, P.R. China.}
\date{}
\maketitle

\begin{abstract}
Carefully analyze influence of the tortoise coordinates $r_*$ and
$t$ on the stable study of the Schwarzschild black hole. Actually,
one should be cautious in using the compact property of the
perturbation field: it is true only with respect with the
coordinate $r$ and proper time or "good time", not the tortoise
coordinates $r_*$ and $t$. Therefore, the mathematical proof used
in reference \cite{wald} is incorrect because of it relying on the
compact property of the perturbation fields. The Schwarzschild
black hole might be unstable\cite{tian1}-\cite{tian5}.

\end{abstract}

The Earth-based interferometers include LIGO, Virgo, GEO600 and
TAMA, and the space-based interferometer is LISA. These
interferometers all aim at the detection of the gravitational
wave. The promising sources of the gravitational wave are the
inspiring binaries  black holes. Among these sources, the
extreme-mass-ratio binaries are primary ones. The
extreme-mass-ratio binary could be modelled by perturbation of a
massive black hole with mass $M$ by a small body with mass $m$.
$\frac m M$ ranges from $10^{-1}$ to $10^{-9}$, though $m$ could
be as large as $100M_{\odot}$. What kind of gravitational fields
could be candidate for the massive black hole? Only those whose
gravitational fields are stable be qualified. Therefore, the
stable study of the black holes' background becomes urgent now.

In general relativity, there are two famous stationary background
metrics for black holes: the Schwarzschild black hole and the Kerr
black hole. The Schwarzschild black hole comes from the massive
spherically symmetric star's complete gravitational collapse,
while the Kerr black hole results from the complete gravitational
collapse of the massive spinning star. Now, the stable properties
of these two black hole are crucial for general relativity
theoretically and astronomically. It is generally believed the
stable problem of the Schwarzschild black hole has been settled,
while the stable properties of the Kerr black hole remains
controversial and unsolved.

We have restudied the stable problem of the Schwarzschild black
hole and found that\textbf{ its stable properties are really
unsolved still}: they really depend on the time coordinate and
where the initial time slice intersects the
horizon\cite{tian1}-\cite{tian4}. In this paper, we carefully
study the problem again and show that the tortoise coordinate
makes us more liable to error on using the compact property of the
perturbation fields. So the generally believed mathematical proof
used in reference \cite{wald} is incorrect because of it relying
on the compact property of the perturbation fields. The stable
properties of the Schwarzschild black-hole still remain unsettled
now, and we only partly infer that it is unstable by comparison
with the similar case in Rindler space time.

Originally, it is in the  Schwarzschild coordinate system where
the stable problem is studied. The well-known perturbation
equation, the Regge-Wheeler equation
\begin{equation}
\frac{\partial^{2}Q}{\partial t^2}-\frac{\partial^{2}Q}{\partial
r^{*2}}+VQ=0,\label{rw}
\end{equation}
is obtained in this very coordinates. In the Regge-Wheeler
equation (\ref{rw}), the tortoise coordinate $r^*$ is defined by
\begin{equation}
r^{*}=r+2m\ln \left(\frac{r}{2m}-1\right)\label{tortoise}
\end{equation}
which approaches $-\infty$ at the horizon $r=2m$ and $\infty$ at
spatial infinity, and the effective potential $V$
\begin{equation}
V=\left(1-\frac{2m}{r}\right)\left[\frac{l\left(l+1\right)}{r^{2}}-\frac{6m}{r^{3}}\right]
\label{potential}
\end{equation}
is positive over $r^*$ from $-\infty$ to $\infty$.

Generally, it is the  Schwarzschild time coordinate $t$ that is
used to define the initial time, and subsequently proved that the
Schwarzschild black hole is stable\cite{vish},\cite{wald}.
Recently, we have discovered that the  Schwarzschild time
coordinate $t$ is not suitable to study the stable problem near
the horizon of the  Schwarzschild black hole where it loses its
meaning actually. If we use  other "good" time coordinates to
define the initial time, the stable properties depend
complicatedly on where the initial time slice intersects with the
black hole: the Schwarzschild black hole is stable when the
initial time slice intersects the future horizon of the black
hole, whereas the Schwarzschild black hole is unstable when the
initial time slice intersects the past horizon of the black
hole\cite{tian1}-\cite{tian4}. These extraordinary results are in
contrast with the conclusion taken as granted that the
Schwarzschild black hole is stable. It makes people puzzled on the
problem.

Later, we study the stable problem of the Rindler space time for
finding the real answer to the Schwarzschild black hole's stable
problem\cite{tian5}. The Rindler space time has the much similar
or almost the same geometrical structure with the Schwarzschild
black-hole, and is mathematically simple for study of the stable
problem. It is found that the same situation exists for the
Rindler space time if we study the scalar field equation
completely in the Rindler space time: its stable properties also
depends on complicatedly on where the initial time slice
intersects with the Rindler space time: the Rindler space time is
stable when the initial time slice intersects its future horizon,
whereas it is unstable when the initial time slice intersects its
the past horizon \cite{tian5}.

Unlike the Schwarzschild black hole, the Rindler space time is one
part of the Minkowski space time, which is the most suitable and
simplest space time to study. We could study the scalar field
equation completely in the Minkowski space time coordinate system
and obtain definite answer to the stable problem of the Rindler
space time: in this very way, it is found that the Rindler space
time is unstable\cite{tian5}. This answer clearly shows which is
correct among the controversial answers to the Rindler space time'
stable problem, that is, the answer whose initial time slice
intersects the past horizon is correct answer\cite{tian5}.

Though we have not study the perturbation equation completely in
Kruskal coordinate system, which is completely too difficult to
study, we still could partly infer and partly guess  that the
Schwarzschild black hole might be unstable by comparison with the
case of the Rindler space time. Nevertheless, there still exists
opposing opinion to our unusual conclusion. Generally, the stable
problem is believed settled by much more mathematically rigorous
proof not based on the above method of the
mode-decomposition\cite{wald}. The Regge-Wheeler equation
(\ref{rw}) could be written as
\begin{equation}
\frac{\partial^{2}Q}{\partial t^2}=\frac{\partial^{2}Q}{\partial
r_*^{2}}-VQ=0,\label{rw2}
\end{equation}
then the operator
\begin{equation}
A=-\frac{d^{2}}{dr_*^{2}}+V,\label{operator}
\end{equation}
is positive and self-conjugate on the Hilbert space $L^2(r_*)$ of
square integral functions of $r_*$ . It is mathematically by these
very properties of the operator $A$ and the compact-supported-ness
of the quantity $Q$ that the Schwarzschild black hole is proven
stable\cite{wald}. The mathematical method is generally believed
correct.

Therefore, if we insist that the Schwarzschild black hole is
unstable, we must answer where the above mathematical proof is
wrong. This is the main purpose of this paper to address.

The main problem in reference \cite{wald} lies on the misleading
concept "compact-supported-ness of the quantity $Q$": this concept
is used in the tortoise coordinates $t$ and $r_*$. Actually, $r_*$
is the spatial tortoise coordinate, which changes the horizon of
the Schwarzschild black hole $r=2m $ finite into negative infinity
by $r_*$. It is well-known the horizon really lies in finite
spatial proper distance from some point $r=r_0>2m$, absolutely not
infinite proper distance. The horizon corresponding to
$r_*\rightarrow -\infty$ is only the effect of the tortoise
coordinate $r_*$. Similarly, the Schwarzschild time coordinate $t$
actually is also a tortoise coordinate. It is also well-known that
a traveller only needs finite proper time to arrive at the horizon
of the Schwarzschild black hole and fall into it. In contrast to
the proper time, it takes the tortoise coordinate $t$ to infinity
for the traveller to arrive at the horizon of the Schwarzschild
black hole. This really means that any finite proper time process
containing the horizon of the Schwarzschild black hole corresponds
to $t\rightarrow \infty$ of the Schwarzschild time coordinate. The
tortoise property of the Schwarzschild time coordinate $t$ makes
it not suitable for the stable study at the horizon: for example,
it may only takes short proper time for some physical fields to
arrive at the horizon, while it corresponds to the Schwarzschild
time coordinate $t\rightarrow \infty$. In this case, the
compact-supported-ness of the physical quantity $Q$ with respect
to the tortoise coordinates $t$ and $r_*$ is really misleading.
 Therefore, the
Reege-Wheeler equation is not suitable for mathematical proof
involving the concept of the compact-supported-ness of the
physical quantities, and the stable proof based on this very
concept is actually false, no matter how it seems mathematically
rigorous. For example, the positive property of the operator $A$
of eq.(\ref{operator}) relies on the compact-supported-ness of the
quantity $Q$; so does the quantity $g(t)=0$ in the following
equation
\begin{equation}
\frac{\partial}{\partial t}\left[\int_{-\infty}^{+\infty}\mid
\dot{Q}\mid
^{2}dr_*+\int_{-\infty}^{+\infty}Q^*AQdr_*\right]=g(t)\label{g(t)}
\end{equation}
where $g(t)$
\begin{equation}
g(t)=-\lim_{R\rightarrow \infty}\left[Q^*\frac{\partial
\dot{Q}}{\partial r_*} -\dot{Q}\frac{\partial Q^*}{\partial r_*}
\right]_{-R}^{+R}.\label{g(t)2}
\end{equation}
When $g(t)=0$, the eq.(\ref{g(t)}) is the eq.(6) in reference
\cite{wald}. As just stated, the real physical process arrives at
the horizon when $t\rightarrow \infty $, though it takes only
finite proper time actually. So, we must study the process at the
tortoise time $t=\infty +\bar{t}$ where $\bar{t}$ is finite.
Similarly, $t$ in the eq.(\ref{g(t)}) is also the same $t=\infty
+\bar{t}$, and the quantity $Q$ is no longer compact-supported in
the tortoise coordinate $r_*$, but in coordinate $r$, that is,
$Q=0$ at $r\rightarrow \infty$ and $Q\ne 0$ at $r=2m$ or
$r_*\rightarrow -\infty$. So, $g(t)$ is generally no longer zero.
The mode-decomposition method is not suitable in the
eq.(\ref{g(t)}), nevertheless, we use the mode-decomposable
solution  $Q_{k}(r_{*})e^{-ikt}$ as an example: while $Q_{k}=B_k
e^{\pm ikr_{*}}$ when $r_{*}\rightarrow \pm \infty$. Actually by
the eq.(\ref{g(t)2}), $g(t)=|B_{k}|^{2}|k|^{2}> 0$ is obtained for
the mode-decomposable solution.

Hence the mathematical proof in reference \cite{wald} fails.

In the following, we use Rindler space time as an example to show
why the compact-supported-ness is not suitable in the tortoise
coordinates. Suppose we solve the scalar field for the Rindler
space time. Because the Rindler space time is one part of the
Minkowski space time, we could write the equation in the Minkowski
coordinates. The Minkowski space time's metric is
\begin{equation}
ds^{2}=-dT^{2}+dZ^{2}+dx^{2}+ dy^{2},\label{orimetric-m}
\end{equation}
and the scalar field equation in the Minkowski space time is
\begin{equation}
-\frac{\partial^{2}\Psi}{\partial T^{2}}+ \frac {\partial ^2
\Psi}{\partial Z^2}+ \frac{\partial^{2}\Psi}{\partial
x^{2}}+\frac{\partial^{2}\Psi}{\partial
y^{2}}-m^{2}\Psi=0.\label{kg in m}
\end{equation}

The Rindler space time corresponds the part $Z>0,\ Z^2-T^2>0$ of
the Minkowski space time. $Z^2-T^2=0$ corresponds the horizon of
the Rindler space time. The boundaries for the eq.(\ref{kg in m})
consist of the horizon $Z^2-T^2=0$ and the spatial infinity
$Z\rightarrow \infty$.

The dependence of $\Psi$ on the variable $x,\ y$ could be
\begin{equation}
\Psi =\psi(Z,T)e^{ik_1x+ik_2y}
\end{equation}
where $\psi$ satisfies
\begin{equation}
\frac{\partial^2\psi}{\partial T^2}= \frac {\partial ^2
\psi}{\partial Z^2}- (k_1^2+k_2^2+m^2)\psi.\label{kg in m2}
\end{equation}
The operator $A_1$ is defined as
\begin{equation}
A_1= -\frac {d^2 }{dZ^2}+ (k_1^2+k_2^2+m^2).\label{operator a1}
\end{equation}

Corresponding the Minkowski coordinates $T$ and $Z$, we could use
the concept of the compact-supported-ness of the $\psi$, that is,
the initial compact data of $\Psi$ will be compact at finite time
$T$. So we could obtain $\psi (T,Z)=0$ as $Z\rightarrow \infty$.
Because the horizon is $Z^2-T^2=0$, generally, $\psi (T,Z)\ne 0$
at $T$. Subsequently, the operator $A_1$ is no longer positive due
to the following fact that
\begin{eqnarray}
I(\psi )&=& \int_{horizon}^{+\infty}\psi^*A_1\psi
dZ=\int_{horizon}^{+\infty} \left[-\psi^*\frac {\partial^2\psi
}{\partial Z^2}+ (k_1^2+k_2^2+m^2)\psi
\psi^*\right]dZ\nonumber \\
&=& \left[-\psi^*\frac {\partial \psi }{\partial
Z}\right]_{horizon}^{\infty}+ \int_{horizon}^{+\infty} \left[\frac
{\partial \psi^* }{dZ}\frac {\partial \psi }{dZ}+
(k_1^2+k_2^2+m^2)\psi \psi^*\right]dZ \label{a3}
\end{eqnarray}
could be negative foe some $\psi$. For example,
$I(B_0(T)e^{-kZ})=\frac
{|B_0(T)|^2}{2k}\left[(k_1^2+k_2^2+m^2)-k^2\right]<0$ for some
$k>\sqrt{k_1^2+k_2^2+m^2}$. This again shows that the Rindler
space time is not stable.

On the other hand, we could write the scalar equation in the
Rindler coordinates. The Rindler metric is
\begin{equation}
ds^{2}=-(1+az)dt^{2}+(1+az)^{-1}dz^{2}+dx^{2}+
dy^{2},\label{orimetric r}
\end{equation}
and the horizon is at $z=-\frac 1a$. The coordinates  $t,\ z$ is
related with $T,\ Z$ by
\begin{equation}
T=\frac2a\sqrt{1+az}\sinh{\frac{at}2},\  \
Z=\frac2a\sqrt{1+az}\cosh{\frac{at}2}.
\end{equation}
The scalar equation in the Rindler coordinates is
\begin{equation}
-\frac1{1+az}\frac{\partial^2\psi}{\partial t^2}+ \frac {\partial
}{\partial z}\left[(1+az)\frac {\partial \psi }{\partial
z}\right]- (k_1^2+k_2^2+m^2)\psi=0.\label{kg in r2}
\end{equation}
Define the spatial tortoise coordinate $z_*$ as
\begin{equation}
z_*=\frac 1a\ln(1+az)
\end{equation}
which makes the horizon $1+az=0$  and the spatial infinity into
$z_*\rightarrow -\infty$ and $z_*\rightarrow +\infty$
respectively. The eq.(\ref{kg in r2}) then becomes
\begin{equation}
\frac{\partial^2\psi}{\partial t^2}- \frac {\partial ^2
\psi}{\partial z_*^2}+(1+az) (k_1^2+k_2^2+m^2)\psi =0.\label{kg in
r3}
\end{equation}
Then the operator $A_2$ is
\begin{equation}
A_2= -\frac {d^2 }{dz_*^2}+
(1+az)(k_1^2+k_2^2+m^2).\label{operator a2}
\end{equation}
The eq.(\ref{kg in r3}) and (\ref{operator a2}) for the scalar
field in the Rindler space time is almost the same as that in the
Schwarzschild black hole (The only difference lie in the spatial
infinity for the effective potential $V=(1+az)(k_1^2+k_2^2+m^2)$
does not fall off to zero in the Rindler space time).

From above argument, the scalar field is not compact in the
tortoise coordinates $t,\ z$ because $\psi$ is not zero at the
"good" finite time $T$ on the horizon $Z^2-T^2=0$. But the
eq.(\ref{kg in r3}) and (\ref{operator a2}) and the Rindler metric
(\ref{orimetric r}) can easily mislead us the wrong concept that
$\psi$ should be compact. Subsequently, we might wrongly prove
that the Rindler space time is stable with respect to the scalar
perturbation.

In the above example, we see it is the tortoise coordinates that
makes us more liable to error and this error may be much more
elusive  to find. This tortoise property of the Schwarzschild time
coordinate $t$ is not noticed by researchers in stable study,
though it is known in other aspects study.

Therefore the stable properties of the Schwarzschild black-hole
still remain unsettled now, and we only partly infer that it is
unstable by comparison with that the Rindler space time. We guess
this conclusion may related with the fact that  stars more or less
are spinning in real world.

\section*{Acknowledgments}

We are supported in part by the national Natural Science
Foundations of China under Grant  No.10475013, No.10373003,
No.10375087, the National Key Basic Research and Development
Programme of China under Grant No 2004CB318000
 and the post-doctor foundation of
China

\end{document}